\RequirePackage{ifpdf}
\documentclass[12pt,letterpaper]{JHEP3}         
\usepackage{amssymb,amsfonts}

\usepackage{epsfig}

\def\be{\begin{eqnarray}}
\def\ee{\end{eqnarray}}
\newcommand{\nn}{\nonumber}
\newcommand\para{\paragraph{}}

\newcommand{\eqn}[1]{(\ref{#1})}

\def\Dslash{\,\,{\raise.15ex\hbox{/}\mkern-12mu D}}
\def\Dbarslash{\,\,{\raise.15ex\hbox{/}\mkern-12mu {\bar D}}}
\def\delslash{\,\,{\raise.15ex\hbox{/}\mkern-9mu \partial}}
\def\delbarslash{\,\,{\raise.15ex\hbox{/}\mkern-9mu {\bar\partial}}}
\def\pslash{\,\,{\raise.15ex\hbox{/}\mkern-9mu p}}
\def\calDslash{\,\,{\raise.15ex\hbox{/}\mkern-12mu {\cal D}}}

\newcommand{\sign}{{\rm sign}}

\def\lae{\mathrel{\mathop{\smash{\lower .5 ex \hbox{$\stackrel<\sim$}}}}}
\def\lae{\mathrel{\mathop{\smash{\lower .5 ex \hbox{$\stackrel>\sim$}}}}}

\usepackage{cite}

\title{Intersecting Branes, Domain Walls  and Superpotentials in 3d Gauge Theories}

\author{Daniele Dorigoni and David Tong\\
Department of Applied Mathematics and Theoretical Physics, \\
University of Cambridge, \\
Cambridge, CB3 0WA, UK\\{\tt d.dorigoni, d.tong@damtp.cam.ac.uk}}

\abstract{We revisit the Hanany-Witten brane construction of 3d gauge theories with ${\cal N}=2$ supersymmetry. Instantons are known to generate a superpotential on the Coulomb branch of the theory. We show that this superpotential can be viewed as arising from the classical scattering of domain wall solitons. The domain walls live on the worldvolume of the fivebranes and their existence relies on the recent observation that the charged  hypermultiplet at the intersection of perpendicular D-branes has non-canonical kinetic terms. We further show how D$p$ branes may be absorbed at the intersection of perpendicular D$(p+4)$-branes where they appear as BPS sigma-model lumps.
}

\begin{document}
\pagestyle{plain} \setcounter{page}{1}
\newcounter{bean}
\baselineskip16pt \setcounter{section}{0}

\section{Introduction and Summary}

It has been known for many years that the quantum dynamics of three-dimensional gauge theories with ${\cal N}=4$ supersymmetry is related to the classical scattering of BPS monopoles \cite{sw,ch,hw,us,fraser,cherkis,ade}. The purpose of this paper is to explain how the quantum dynamics of three-dimensional gauge theories with ${\cal N}=2$ supersymmetry is related to the classical scattering of BPS domain walls. At the same time, we will shed light on how superpotentials arise in certain configurations of intersecting branes.

\subsubsection*{A Review of Branes and Magnetic Monopoles}

\begin{figure}[!h]
\begin{center}
\includegraphics[height=1.6in]{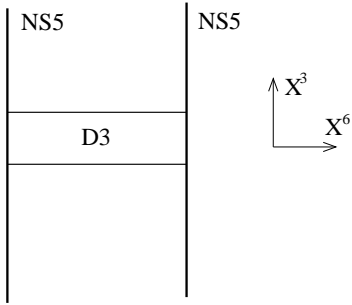}
\end{center}
\caption{In the 1990s, you could solve supersymmetric gauge theories by drawing diagrams like this.}\label{ifig}
\end{figure}

The Hanany-Witten brane construction  provides the simplest method to see the relationship between three-dimensional gauge theories and monopole moduli spaces \cite{hw}. $N$ D3-branes are suspended between a pair of NS5-branes as shown in Figure 1. Their worldvolumes span:
\be D3&:& \ \ 0126 \nn\\ NS5&:&\ \ 012345\nn\ee
The theory on the D3-branes is $d=2+1$-dimensional $U(N)$ gauge theory with ${\cal N}=4$ supersymmetry. Each D3-brane is free to move in the $X^3$, $X^4$ and $X^5$ directions. These correspond to the expectation values of the three adjoint scalars in the vector multiplet. For a generic configuration, the $U(N)$ gauge symmetry is broken to $U(1)^N$ and each of these $N$ photons can be dualised to a  periodic scalar.  
The result is a $4N$-dimensional configuration space;  this is the Coulomb branch of the gauge theory. The low-energy dynamics of the gauge theory is governed  by the metric on the Coulomb branch.

\para
The metric on the Coulomb branch can be determined by taking a different perspective on the brane picture. We start by  performing an S-duality so that the NS5-branes are replaced by D5-branes. The theory on the D5-branes is $SU(2)$ Yang-Mills in $d=5+1$ dimensions. The D3-branes stretched between them appear as $N$ magnetic monopoles \cite{diac}. This strongly suggests that the quantum corrected metric on the Coulomb branch of the 3d  $SU(N)$ gauge theory coincides with the metric on the classical moduli space of $N$ monopoles. This correspondence has been confirmed by a number of explicit calculations in the field theory \cite{sw,ch,us,fraser}. 


\subsubsection*{Branes and Superpotentials}

There is a simple modification of the brane set-up which realises $d=2+1$-dimensional gauge theories with ${\cal N}=2$ supersymmetry on the worldvolume of the D3-branes. We need only rotate one of the NS5-branes \cite{brane1,brane2,brane3}. After this rotation, it is usually referred to as an NS5$'$-brane. The worldvolume directions now span:
\be D3&:& \ \ 0126 \nn\\ NS5&:&\ \ 012345 \nn\\ NS5'&:&\ \ 012378\nn\ee
The D3-branes are  free to move only in the $X^3$ direction. At a generic point in the classical moduli space, we may again dualise the $N$ photons, leaving ourselves with $2N$ low-energy degrees of freedom. We would like to know the low-energy dynamics of these modes.

\para
This time, the brane configuration does not provide much of a hint. Instead, it is simpler to turn to an explicit analysis of the 3d gauge theory. This was first done for $U(2)$ gauge theories by Affleck, Harvey and Witten \cite{ahw}.  The centre of mass motion of the branes is free, leaving us with just two interacting scalars: the separation of the branes $X^3$ and the dual photon $\sigma$.  These combine to form the lowest components of a chiral multiplet: $Y = X^3+i\sigma$. The  result of \cite{ahw} (see also \cite{dbho,ahiss}) is that there is an instanton-induced superpotential for this field
\be {\cal W} \sim e^{-Y}\label{w}\ee
This means that the configuration of two D3-branes is unstable and the branes repel each other. Similar behaviour occurs for $N\geq 3$ D3-branes, where adjacent D3-branes are mutually repulsive.

\para
It is natural to ask how we can see this result directly from the brane picture. The discussion of ${\cal N}=4$ theories above suggests an obvious strategy: perform an S-duality, write down the theory on the D5-branes, and identify the D3-branes as an appropriate soliton configuration. The dynamics of these solitons should reproduce the superpotential ${\cal W}$.


\para
The trouble with implementing this strategy in the past was that the theory on non-parallel D5-branes did not seem to admit any soliton solutions which could be identified with the D3-branes. A string stretched between a D5 and D5$'$-brane has 4 Dirchlet-Neumann directions and gives rise to a familiar hypermultiplet, albeit coupled in a non-familiar manner which preserves only $d=3+1$-dimensional Lorentz invariance on the mutual D5-brane worldvolumes \cite{neil}. 
Recently, however, this system was revisited by Mintun, Polchinski and Sun \cite{mps} where they argued that the  hypermultiplet fields should be endowed with non-canonical kinetic terms. As we review below, this introduces solitons into the D5-brane theory which can be identified with the D3-branes. These solitons are kinks. In this paper we compute the force between two kinks and show that it indeed reproduces the quantum superpotential \eqn{w}. 

\para
Most of this paper is devoted to telling the story above.  We start, in Section \ref{intbsec}, by reviewing the proposal of \cite{mps} for the non-canonical kinetic terms. In Section \ref{kinksec} we describe some properties of the resulting kink solutions and how they appear as magnetic monopoles in each of the D5-branes. In Section \ref{multikink}, we compute the force between kinks and show that it reproduces the expected field theory result \eqn{w}.  We also  include an appendix which  provides an alternative description of the hypermultiplet degrees of freedom in terms of a gauged linear model.

\subsubsection*{Branes, Instantons and Lumps}

To end the paper, we turn to  a different consequence of the proposal of \cite{mps}. Consider the following system of D-branes:
\be D1&:& \ \ 01 \nn\\ D5&:&\ \ 012345 \nn\\ D5'&:&\ \ 012378\nn\ee
If there are multiple D5-branes of either kind, then it is well known that the D1-strings can be absorbed into their worldvolume where they appears as an instantons \cite{douglas1}. But what if there is just a single D5 and a single D5${}^\prime$ brane? There is no gauge symmetry enhancement at the intersection, so the D1-branes cannot appear as an instantons. Yet an analysis of the D1-brane worldvolume theory shows that it now enjoys two Higgs branches, which tells us that  D-strings can absorbed in two different ways in the region where the D5-branes overlap.

\para
In section 5, we show how this can be understood through the proposal of \cite{mps}. The theory on the worldvolume of intersecting D5-branes admits a class of BPS solitons known as sigma-model lumps in which the worldvolume wraps the hypermultiplet target space. In fact, because of the structure of the target space, we will see that two different BPS solitons exist. We show that the moduli spaces of these two types of solitons coincide with the two Higgs branches of the D1-brane gauge theory.

\section{Intersecting Branes}\label{intbsec}

In this section we  review  the proposal of \cite{mps} for the worldvolume theory on the D5 and D5$'$ branes. Because this system preserves only $d=3+1$ dimensional Lorentz invariance, we must decompose the fields in a slightly unusual manner. Our notation will be
\begin{itemize}
\item D5-brane: We decompose the worldvolume into $x^\mu$, $\mu=0,1,2,3$ and $z=x^4+ix^5$. Correspondingly, we decompose the gauge field as $A_\mu$ and $A_z=A_4-iA_5$. There are two complex scalars: $W=X^7+iX^8$ and $\Phi=X^6+iX^9$.
\item D5$'$ brane: We decompose the worldvolume into $x^\mu$, $\mu=0,1,2,3$ and $w=x^7+ix^8$. Correspondingly, we decompose the gauge field as $A'_\mu$ and $A'_w=A_7-iA_8$. There are two complex scalars: $Z=X^4+iX^5$ and $\Phi'=X^6+iX^9$.
\end{itemize}
Quantisation of a fundamental string stretched between the D5 and D5$'$ branes yields a hypermultiplet; we will describe these fields in more detail below. 

\para
The worldvolume theory of this system was first discussed in \cite{neil}. The bosonic Lagrangian  takes the form
\be S &=& -\frac{1}{2g^2}\int d^4x dzd\bar{z}\ \Bigg\{\frac{1}{2}F_{\mu\nu}F^{\mu\nu} + \left( F_{45}- D\,\delta^2(z,\bar{z}) \right)^2 
+|F_{\mu z}|^2+|\partial_\mu W|^2 \nn\\ &&\ \ \ \ \ \ \ \ \ \ \ \ \ \ \ \ \ \ \ \ \ \ \ \ \ \ \ \ \ \ \ \ + |\partial_z W - F\delta^2(z,\bar{z})|^2 + |\partial_\mu\Phi|^2 + |\partial_z \Phi|^2  \Bigg\}\nn\\
 \nn\\ && -\ \frac{1}{2g^2}\int d^4x dwd\bar{w}\ \Bigg\{\frac{1}{2}F'_{\mu\nu}F'^{\mu\nu} + \left( F'_{78}+ D\,\delta^2(w,\bar{w}) \right)^2 
+|F'_{\mu w}|^2+|\partial_\mu Z|^2 \nn\\ &&\ \ \ \ \ \ \ \ \ \ \ \ \ \ \ \ \ \ \ \ \ \ \ \ \ \ \ \ \ \ \ \ + |\partial_w Z - F\delta^2(w,\bar{w})|^2 + |\partial_\mu\Phi'|^2 + |\partial_w \Phi'|^2  \Bigg\} \nn\\ && -\ \frac{1}{g^2}\int d^4x\ {\cal L}_{\rm hyper}\label{lag}\ee
Here $D$ and $F$ are the familiar D-terms and F-terms associated to the hypermultiplet. As we now explain, the explicit expressions for them depend on our choice for the hypermultiplet action, ${\cal L}_{\rm hyper}$. 

\subsubsection*{Canonical Kinetic Terms}

The simplest assumption is that the hypermultiplet kinetic terms take the canonical form \cite{neil}. In this case, the hypermultiplet consists of two complex scalar fields, $q$ and $\tilde{q}$ and the bosonic Lagrangian is given by
\be {\cal L}_{\rm hyper} = |{\cal D}_\mu q|^2 + |{\cal D}_\mu \tilde{q}|^2 + |\Phi-\Phi'|^2(|q|^2 + |\tilde{q}|^2)\label{notright}\ee
Here ${\cal D}_\mu q = \partial_\mu q- iA_\mu q  + iA'_\mu q$, as appropriate for a field with charge $(+1,-1)$ under the $U(1)$ gauge groups on the two D5-branes. Similarly, $\tilde{q}$ has charge $(-1,+1)$ and covariant derivative  ${\cal D}_\mu \tilde{q} = \partial_\mu \tilde{q}+ iA_\mu \tilde{q}  - iA'_\mu \tilde{q}$. The final term in \eqn{notright} reflects the fact that these fields become massive if the D5-branes are separated in the $X^6$ or $X^9$ directions. The corresponding $D$ and $F$-terms in \eqn{lag} are
\be D = \frac{1}{2}\left(|q|^2- |\tilde{q}|^2\right)\ \ \ \ {\rm and}\ \ \ \ F = \frac{\tilde{q}q}{\sqrt{2}}\nn\ee
These can be thought of as the moment maps for the $U(1)$ gauge action.

\para
As explained in \cite{mps}, there is a problem with the action \eqn{notright}; when the D5-branes are separated, it does not admit any soliton solutions. Instead, the unique vacuum state is $q=\tilde{q}=0$. This means that if the worldvolume theory of the D5-branes is given by \eqn{lag} and \eqn{notright} then there is no candidate field configuration for the D3-branes.  It was proposed in \cite{mps} that the resolution to this puzzle is that the hypermultiplet fields have different kinetic terms.

\subsubsection*{Non-Canonical Kinetic Terms}

There are various restrictions on the form that the kinetic terms in ${\cal L}_{\rm hyper}$ can take. First, the requirement of 8 supercharges restricts the target space to have a hyperK\"ahler metric. Next, the fact that we want to couple the hypermultiplet to a $U(1)$ gauge field means that the metric should have a (tri-holomorphic) $U(1)$ isometry. Finally, the metric should have one further $U(1)_R$ isometry which leaves one of the three complex structures invariant (and rotates the other two). This ensures that the field theory has a $U(1)_R$ R-symmetry, a property which can be traced to the  $U(1)_{45}\times U(1)_{78}$ rotational symmetry of the brane configuration.  

\para
There is a well-known metric which obeys all these requirements. It is known as the  Gibbons-Hawking metric \cite{gh} and takes the form
\be ds^2 =  V(\vec{r}) \,d\vec{r}\cdot d\vec{r} + V^{-1}(\vec{r})\left(d\theta + \vec{\omega}\cdot d\vec{r}\right)^2\label{metric}\ee
The metric is parameterised by the three-vector $\vec{r}=(r^1,r^2,r^3)$ and the periodic coordinate $\theta\in [0,4\pi)$. The connection $\vec{\omega}$ is given by $\nabla\times\vec{\omega}=\nabla V$ and  the function $V(\vec{r})$ takes the form
\be V(\vec{r}) = c + \sum_n\frac{1}{|\vec{r}-\vec{\kappa}_n|}\label{v}\ee
for some constant $c$ and choice of centres $\vec{\kappa}_n$; we'll have more to say about these parameters shortly. 

\para
The tri-holomorphic isometry is associated to shifts of $\theta$. This is the symmetry that we gauge. The hypermultiplet Lagrangian is given by
\be {\cal L}_{\rm hyper} =\frac{1}{2}g_{ab}(\vec{r})\,{\cal D}_\mu r^a\,{\cal D}^\mu r^b +2|\Phi-\Phi'|^2V^{-1}(\vec{r})\,\label{lhyper}\ee
where we've defined $r^4\equiv \theta$ and the covariant derivatives are given by ${\cal D}_\mu \vec{r}=\partial_\mu \vec{r}$ and ${\cal D}_\mu\theta  = \partial_\mu \theta + 2A_\mu - 2A'_\mu$. Note that the shift charges of $\theta$ are $(2,-2)$ due to its $4\pi$ periodicity. The scalar fields $\Phi-\Phi'$ now couples to the length-squared of the Killing vector $k_\theta = 2\partial_\theta$  \cite{agf}. (The factor of 2 in front of the potential can be traced to the shift charge of $\theta$).

\para
The metric has the desired $U(1)_R$ isometry if all the centres are colinear, so that $\vec{\kappa}_n = (0,0,\kappa_n)$. The complex coordinate $r^1+ir^2 = \rho e^{i\alpha}$ then has charge $+2$ under  $U(1)_R$. (Note: when all the centres coincide, the metric has an enhanced $SU(2)_R$ symmetry, under which the bosonic coordinate transform in the ${\bf 3}$ while the fermions transform in the ${\bf 2}$. Correspondingly, the bosonic coordinate carries charge $+2$ under $U(1)_R$ while the fermions carry charge $+1$). 

\para
The $D$ and $F$-terms  are moment maps associated to the tri-holomorphic action. When the centres are colinear, $\vec{\kappa_n} = (0,0,\kappa_n)$, both are linear in $\vec{r}$. However, rather surprisingly, the $D$-term is only piecewise continuous: 
\be D = 2(r^3-\kappa_n)\ \ \ {\rm for}\ r^3\in [\kappa_n,\kappa_{n+1})\ \ \ {\rm and}\ \ \ \ F = \frac{r^1+ir^2}{\sqrt{2}}\label{dandf}\ee
The D-term jumps at each of the centres. Although such behaviour is unusual in a field theory, it follows simply because the coordinates introduced in \eqn{metric} are not well-suited to cover the whole manifold. 
In Appendix \ref{glsmapp}, we provide a different description in which the D-term is manifestly continuous throughout the manifold.

\para
 The proposal of \cite{mps} is to take an infinite string of centres, with spacing
\be \vec{\kappa}_n = (0,0,2\pi n)\ \ \ \ \ n\in {\bf Z}\label{centres}\ee
This proposal is not without its difficulties. 
Summing over an infinite number of centres in \eqn{v} gives rise to a logarithmic divergence. This can, in part, be accommodated by a suitable shift of the constant $c$. For example, if we choose to sum over the integers $n\in[-N,+N]$, then the log divergence can be absorbed by the shift $c\rightarrow c - (\log N)/\pi$. However, the divergence now rears its head if we move too far in the $\rho^2=(r^1)^2 + (r^2)^2$ direction, where the metric is no longer positive definite. This was interpreted in \cite{mps} as a breakdown of the low-energy effective description, where more information from the underlying string theory is needed. 

\para
In this paper, we shall work with the regularised metric in which we sum over only a finite number of centres: $n\in[-N,+N]$. However, the results we shall derive will be independent of $N$.

\para 
The most significant difference between the canonical \eqn{notright} and non-canonical \eqn{lhyper} kinetic terms lies in the vacuum structure. The requirement that the $D$ and $F$ terms vanish means that there are multiple classical ground states, given by
\be \vec{r}= \vec{\kappa}_n\ \ \ \ \ n\in {\bf Z}\label{ground}\ee
How to interpret these multiple ground states? It was argued in \cite{mps} that they should be thought of as physically equivalent. That is, we should quotient the hypermultplet target space by a freely acting discrete symmetry which identifies these different vacua.  Of course, such a quotient only really make sense in the strict $N\rightarrow \infty$ limit so is rather tricky. Explicit expressions for the local quotient were presented in \cite{mps}, albeit in different coordinates from those used here.

\section{Kinks as Magnetic Monopoles}\label{kinksec}

The existence of  multiple ground state in the covering space has an important consequence: it leads to the existence of different kinds of soliton solutions in the theory.  In this section we discuss domain walls, or kinks. In Section \ref{d1d5sec}, we discuss sigma-model lumps. 

\para
Start by separating the D5 and D5$'$-branes by a distance $v\alpha'$ in the $X^6$ direction; this means that we set ${\rm Re}\,\langle \Phi-\Phi'\rangle = v$. The D5-branes remain coincident in the $X^9$ direction, with ${\rm Im}\,\langle\Phi\rangle = {\rm Im}\,\langle\Phi'\rangle=0$. The vacuum states \eqn{ground} persist when we separate the D5-branes because $V(\vec{r}=\vec{\kappa}_n)^{-1}=0$ so the potential energy in \eqn{lhyper} vanishes. This allows for the possibility of kink solutions interpolating between different vacua $\vec{r}=\vec{\kappa}_n$. These kinks were identified with the D3-branes stretched between D5-branes in \cite{mps}. We now review a number of properties of these solitons.

\para
In the brane configuration described in the introduction, the D3-branes are localised in the $X^3$ direction. For this reason we look for kinks which interpolate between different vacua as $x^3\rightarrow \pm \infty$. 
From the perspective of the D5-brane, the kink is a co-dimension 3 object lying at a point in ${\bf R}^3$ parameterised by $x^i$ with $i={3,4,5}$. The kink sources the magnetic fields $F_{ij}$ on the D5-brane, where 
 it appears as a magnetic monopole. However, there is a subtlety in how we define the gauge field that carries the magnetic charge. In particular, the original gauge field $F_{ij}$ obeys the Bianchi identity $\epsilon_{ijk}\partial_iF_{jk}=0$ everywhere in space and hence  cannot carry a magnetic charge. Instead, we define 
\be {\cal F}_{45} = F_{45} - D\,\delta^2(z,\bar{z})\label{newf}\ee       
while ${\cal F}_{\mu\nu}=F_{\mu\nu}$ for $\mu,\nu\neq 4,5$. Naively it looks as if ${\cal F}_{45}$ suffers from the delta-function singularity at $z=0$. But, as we will show in more detail shortly, ${\cal F}_{45}$ turns out to be  asymptotically smooth; the explicit delta function is cancelled by a corresponding term in $F_{45}$. For now, we note that it is ${\cal F}_{\mu\nu}$ which is the  field strength which appears in the Lagrangian \eqn{lag} in the standard Maxwell form
${\cal F}_{\mu\nu}{\cal F}^{\mu\nu}$. 

\para
 The field strength ${\cal F}_{\mu\nu}$ is not constrained to obey the Bianchi identity; instead $\epsilon_{ijk} \partial_i {\cal F}_{jk} = -2\partial_3 D\,\delta^{2}(z,\bar{z})$. 
This ensures that if we  integrate the associated magnetic field ${\cal B}_i=\frac{1}{2}\epsilon_{ijk}{\cal F}_{jk}$ over an ${\bf S}^2$ surrounding the kink, we find the magnetic charge
\be \int_{{\bf S}^2}dS_i\,{\cal B}_i= \left[\int_{{\bf R}^2_+}-\int_{{\bf R}^2_-} \right]dzd\bar{z}\ {\cal B}_3 =-\Delta D\nn\ee
%
%
%
where, in the first equality, we have deformed the ${\bf S}^2$ into two planes at $x_3=\pm \infty$. 
With the centres given by \eqn{centres},  the kink interpolating from vacuum $\kappa_n$ to vacuum  $\kappa_{n+1}$, has magnetic charge is
\be \int_{{\bf S}^2}dS_i\,{\cal B}_i = -4\pi\nn\ee
which is consistent with Dirac quantisation. 

\para
There is a similar story for the magnetic field on the D5$'$-brane. Now the appropriate magnetic field is defined as
\be {\cal B}'_3 =  {\cal F}_{78}' =F'_{78} + D\,\delta^2(w,\bar{w})\nn\ee
while ${\cal F}'_{\mu\nu} = F'_{\mu\nu}$ for $\mu,\nu\neq 7,8$. Integrating over the appropriate 2-sphere, a single kink has magnetic charge $
 \int_{{\bf S}^2} {\cal B}' = +4\pi$.

\subsubsection*{BPS Equations}

BPS kinks in hyperK\"ahler, non-linear sigma models were first studied in \cite{abrahams} and further explored in \cite{pauljerome}. The new ingredient here is the coupling to the fields on the D5-branes, under which the kink carries magnetic charge. 

\para
The BPS kinks have $r^1=r^2=0$. In what follows, we write $r^3\equiv r$; this depends only on $x^3$. Meanwhile, the fields $\Phi$ and $A_{3,4,5}$ on the D5-brane will depend on $x^3$, $x^4$ and $x^5$; the fields $\Phi'$ and $A'_{3,7,8}$ on the D5$'$-brane will depend on $x^3$, $x^7$ and $x^8$. All other fields are zero. 
We can derive the first order kink equations by writing the tension $T$ of the kink thus: 
\be T &=& \frac{1}{g^2} \int dx^3 dz d\bar{z}\ \Bigg\{ \frac{1}{2}\left(\partial_3\Phi + (F_{45} - D\delta^2(z,\bar{z})\right)^2 + D\,\partial_3\Phi\,\delta^2(z,\bar{z}) \nn\\ &&\ \ \ \ \ \ \ \ \ \ \ \ \ \ \ \ \ \ \ \ \ \ \ \ \ \ \ \ 
+\ \frac{1}{2} |\partial_z\Phi -i F_{3z}|^2 -\frac{1}{2}\epsilon_{ijk}F_{ij}\partial_k\Phi\Bigg\}
\nn\\ && +\ \frac{1}{g^2}\int dx^3 dw d\bar{w}\ \Bigg\{ \frac{1}{2}\left(\partial_3\Phi' + (F'_{78} + D\delta^2(w,\bar{w})\right)^2  -D\,\partial_3\Phi'\,\delta^2(w,\bar{w})
\nn\\ &&\ \ \ \ \ \ \ \ \ \ \ \ \ \ \ \ \ \ \ \ \ \ \ \ \ \ \ \ \ +\ \frac{1}{2}|\partial_w\Phi' -i F'_{3w}|^2 - \frac{1}{2}\epsilon_{ijk}F'_{ij}\partial_k\Phi'\Bigg\}  \nn\\
&&  +\  \frac{1}{g^2} \int dx^3 \ \Bigg\{\frac{1}{2}V(r) \left(\partial_3 r -2 (\Phi-\Phi') V^{-1}(r)\right)^2 +2 (\Phi-\Phi')\partial_3r  \nn\\ 
&&\ \ \ \ \ \ \ \ \ \ \ \ \ \ \ \ \ \ \ \ \ \ \ \ \ \ \ \ 
 +\ \frac{1}{2}V^{-1}(r)|{\cal D}_3\theta + \vec{\omega}\cdot \partial_3\vec{r}|^2\Bigg\}\nn\ee
Here $x^i$ refers to $x^{3,4,5}$ on the D5-brane and $x^{3,7,8}$ on the D5$'$-brane. 
The Bogomolnyi equations can be found sitting within the total squares. They are the domain wall equation
\be \partial_3 r = 2(\Phi-\Phi')V^{-1}(r)\ \ \ {\rm and}\ \ \ {\cal D}_3\theta + \vec{\omega}\cdot\partial_3\vec{r}=0 \label{dw}\ee
together with the  BPS monopole equations
\be {\cal B}_i = -\partial_i \Phi\ \ \ {\rm and}\ \ \ {\cal B}_i'=- \partial_i\Phi'\label{mono}\ee
%
%
%
we remind the reader that the  definition \eqn{newf} of ${\cal B}_i = \frac{1}{2}\epsilon_{ijk}{\cal F}_{jk}$ involves the scalar field $r$. 
This set of equations are related, but seemingly not identical, to those given in \cite{mps}.  One can check that solutions to these first order equations are also solutions to the full, second-order equations of motion.

\para
Using the fact that the D-term is linear in $r$ \eqn{dandf}, the tension of any domain wall satisfying these first order equations is given by
\be
T&=& 
 \frac{2v}{g^2}\Delta r - \frac{1}{g^2}\int_{{\bf S}^3} \Phi \vec{B}\cdot d\vec{S} -  \frac{1}{g^2}\int_{{\bf S'}^3} \Phi' \vec{B'}\cdot d\vec{S}
 \nn\\ &=& 
 \frac{2v}{g^2}\Delta r  = \frac{4\pi v}{g^2}\nn\ee
 where the last two terms vanished because $\Phi$ and $\Phi'$ are constant asymptotically while, as described above, $\vec{B}$ and $\vec{B}'$ obey the Bianchi identities and carry no magnetic charge. This is the same as the tension of a  magnetic monopole, corresponding to a D3-brane stretched between parallel D5-branes.

\para
The interplay between kinks and monopoles in these BPS equations has some interesting antecedents. It was noticed long ago that kinks in non-linear sigma models, obeying very similar equations to \eqn{dw}, share a number of features with magnetic monopoles \cite{abrahams}. One explanation for this was provided in \cite{monohiggs}; when gauge  theories lie in the Higgs phase, monopoles can be viewed as kinks on vortex flux tubes. Here we see that there is a second, close relationship between kinks and magnetic monopoles.

\subsubsection*{Some Properties of the Solution}

We have not found explicit solutions for the coupled equations \eqn{dw} and \eqn{mono}. However, it is a simple matter to get a handle on the asymptotic properties. Using the Bianchi identity for the field strength $F_{ij}$, we have
\be \nabla^2 \Phi = 2\partial_3 r\,\delta^2(z,\bar{z})\nn\ee
and similar for $\Phi'$. The profile of the scalar field is therefore given by
\be
 \Phi(x_3,x_4,x_5) = -\frac{1}{2\pi}\int_{-\infty}^\infty dx \frac{\partial_x r(x)}{\sqrt{(x_3-x)^2+x_4^2+x_5^2}}+\Phi_0 \label{profile}
 \nn\ee
with $\Phi_0$ the asymptotic value. 

\para
This profile has interesting behaviour. As we approach the $x^4=x^5=0$ line, the scalar field looks like $\Phi \sim \log (x_4^2+x_5^2)$, with a coefficient that depends on $\partial_3 r(x^3)$. One might have hoped that in the centre of the kink, we would have $\Phi-\Phi'\rightarrow 0$, reflecting the meeting of the two D5-branes. (Analogous  behaviour is seen, for example, in the centre of a 't Hooft Polyakov monopole). However, this is not the case. The two fields are related by $\Phi'(x^3,x^7,x^8) = - \Phi(x^3,x^4,x^5)$, meaning that both diverge logarithmically in opposite directions. This is telling us that the  low-energy effective theory \eqn{lag} cannot be trusted at small distances.

\para
Our real interest in this paper is in the long distance behaviour of the theory. When  $x^4$, $x^5$ are large, $\Phi\sim \Phi_0 + \Delta r/ |\vec{x}|$.   It's instructive to use the profile \eqn{profile} to compute the magnetic charge carried by ${\cal B}_i$. The two are related through the Bogomolnyi equation \eqn{mono}. The magnetic charge is
\be \int_{{\bf S}^2}dS_i\,{\cal B}_i= \left[\int_{{\bf R}^2_+}-\int_{{\bf R}^2_-} \right]dzd\bar{z}\ {\cal B}_3 = \left[\int_{{\bf R}^2_+}-\int_{{\bf R}^2_-} \right]dzd\bar{z}\ \left(-\partial_3\Phi\right)  \nn\ee
From the integral expression \eqn{profile}, we have
\be
\int dzd\bar{z}\ \partial_3\Phi &=& -\frac{1}{2\pi} \int dx dx_4dx_5\ \frac{(x-x_3) \partial_x r(x) }{[(x_3-x)^2+x_4^2+x_5^2]^{3/2}} \nn\\ &=&
 -\int dx\ \partial_x r(x)\, \sign(x-x_3)\nn\ee
%
%
%
%
This can be happily evaluated in the limit  $x_3 \to \pm \infty$ to get
\be
 \int_{{\bf S}^2}dS_i\,{\cal B}_i=\left[\int_{{\bf R}^2_+}-\int_{{\bf R}^2_-} \right]dzd\bar{z} \ \left(-\partial_3\Phi\right) = -\Delta r - \left( \Delta r \right) = -2\Delta r 
\nn\ee%
This computations shows that the magnetic charge of ${\cal B}_i$ comes from the $1/|\vec{x}|^2$ fall-off; there is no extra contribution from a delta-function at $x^4=x^5=0$. This is because, although $\Phi$ is log-divergent here, the coefficient of the log is proportional to $\partial r(x^3)$ and vanishes asymptotically. 

\para
In contrast, we can look at the magnetic charge carried by the original  field strength $B_i$. From \eqn{newf}, this is given by 
\be \int_{{\bf S}^2}dS_i\, B_i=  \left[\int_{{\bf R}^2_+}-\int_{{\bf R}^2_-} \right]dzd\bar{z}\ \left(-\partial_3\Phi + 2r\delta^2(z,\bar{z})\right) =0 \nn\ee
 We see that, in this case, the explicit contribution from the delta-function cancels the $1/|\vec{x}|^2$ contribution from $\partial \Phi$ that we just computed and $\vec{B}$ carries vanishing magnetic charge as we earlier anticipated.

\section{Kink Dynamics}\label{multikink}

We turn now to the main purpose of this paper. We compute the classical low-energy effective theory of the kinks and show that this coincides with the effective quantum dynamics of the $d=2+1$ dimensional gauge theory. 

\subsection{Dynamics of a Single Kink}

We start by describing the low-energy dynamics of a single kink. There are two collective coordinates. The first is simply the centre of mass, $X$, of the kink in the $x^3$ direction.

\para
The second collective coordinate is an internal mode first identified in \cite{abrahams}. It arises from acting on the kink solution with the $U(1)_F$ flavour symmetry. In the present case, this flavour symmetry is gauged on the two D5-branes. Nonetheless, the collective coordinate survives as  a global gauge transformation. This is very similar to the way the extra internal collective coordinate of a 't Hooft-Polyakov monopole arises. 

\para
Roughly speaking, this internal collective coordinate can be thought of as the value of $\theta$ of the kink. However there is a subtlety in this definition that we now spell out because it will be important in what follows. The relevant equation is from \eqn{dw}. We are free to work in the gauge $A_3=A_3'=0$, where the second BPS equation becomes
\be \partial_3\theta + \vec{\omega}\cdot\partial_3\vec{r}=0\label{betterbog}\ee
Recall that $\vec{\omega}$ is defined by $\nabla\times \vec{\omega} = \nabla V$ with $V$ given in \eqn{v}. If we write $r^1+ir^2 = \rho e^{i\alpha}$ and, as before $r^3=r$, then solving for $\vec{\omega}$ gives
\be
\vec{\omega}\cdot\partial_3\vec{r}=\sum_n
\frac{r-\kappa_n}{\sqrt{\rho^2+(r-\kappa_n)^2}}\partial_3\alpha
\nn\ee
The BPS kinks lie along the line $\rho=0$. But we see that, like the complex phase of a cheshire cat, the dynamics of $\alpha$ does not disappear when $\rho=0$. This, of course, is due to our attempt to cover a topologically non-trivial manifold with a single set of coordinates in \eqn{metric}.

\para
For our purposes, it means that after setting $\rho=0$, the Bogomolnyi equation \eqn{betterbog} can be written as 
\be \partial_3 \sigma = 0\ \ \ \ {\rm with}\  \ \ \ \ \sigma=\theta+ Q\alpha\label{sigma}\ee
 where $Q=\sum_n\sign(r-\kappa_n)$ counts the number of centres to the left of the kink minus the number of centres to the right. The value of $\sigma$ provides the second collective coordinate. (This difference between $\sigma$ and $\theta$ will not be important in the discussion of a single kink; it will however, be crucial when we come to discuss the dynamics of a pair of kinks). 

\para
Having identified the two collective coordinates, we can proceed to write down a low-energy effective action for the kink. We work in the moduli space approximation, in which both collective coordinates are allowed to vary slowly: $X=X(x^\mu)$ and $\sigma=\sigma(x^\mu)$ with $\mu=0,1,2$ spanning the worldvolume of the kink. Substituting this ansatz into the original action \eqn{lag} yields the expression for the low-energy kink dynamics
\be S_{\rm kink} = \int d^3x\ \frac{T}{2}\partial X^2 + \frac{T}{2v^2}\,\partial\sigma^2 = \int d^3x\ \frac{T}{2v^2}\,\partial \bar{Z}\partial Z\label{onekink}\ee
where $T=4\pi v/g^2$ is the tension of the kink and, in the second equality, we have introduced the holomorphic combination $Z = vX + \sigma$.  The fermionic zero modes complete this action into one with ${\cal N}=2$ supersymmetry, with $Z$ the lowest component of a chiral superfield.

\para
We now proceed to massage this action into that of a $U(1)$ gauge theory. First, we rescale the centre of mass collective coordinate and define $\phi = v^2 X$. Next, we dualise the periodic scalar in favour of an Abelian field strength, $f_{\mu\nu} = v \epsilon_{\mu\nu\rho}\partial^\rho \sigma$. The end result is the bosonic part of a ${\cal N}=2$ $U(1)$ gauge theory,
\be S_{\rm kink} = \int d^3x\ \frac{1}{4e^2} f_{\mu\nu}f^{\nu\nu}  + \frac{1}{2e^2}\partial \phi^2\nn\ee
where $e^2 = 4\pi/g^2 v^4$.

 \subsection{Supersymmetric Dynamics of a Pair of Kinks}

We now want to study the low-energy dynamics of a pair of kinks. In this case, we will not dualise to find a $U(2)$ gauge theory. Instead,   will show that the classical dynamics of the kinks correctly captures the quantum dynamics of the non-Abelian $U(2)$ gauge theory. 

\para
The key observation is that there are no solutions to the BPS equations \eqn{dw} for domain walls interpolating between non-adjacent vacua. If we build a field configuration consisting of two, far-separated, domain walls then these walls will experience a repulsive force. We will compute the repulsive force between domain walls in two different ways. We start, here, by giving an argument based on symmetries alone. We subsequently present a more direct computation of the force. 

\para
Our first approach will be to write down the most general effective action for the pair of kinks, consistent with the symmetries of the theory. These symmetries include the four supercharges that are preserved by the kink solution, as well as the $U(1)_R$ symmetry of the underlying theory. Taken individually, each kink has two collective coordinates which we call $X^{(1)}$, $\sigma^{(1)}$ and $X^{(2)}$, $\sigma^{(2)}$. We decompose them into holomorphic coordinates corresponding to the centre of mass, $Z$ and relative separation $Y$
\be Z &=& \frac{v}{2}(X^{(1)}+X^{(2)}) + \frac{1}{2}(\sigma^{(1)} + \sigma^{(2)})\nn\\ Y &=& \frac{v}{2}(X^{(1)}-X^{(2)}) + \frac{1}{2}(\sigma^{(1)} - \sigma^{(2)})\label{y}\ee
Each of these is the lowest component of a chiral superfield. 
The collective coordinate $Z$ contains Goldstone modes and is free. The would-be collective coordinate  $Y$ is more interesting. Since we know that no two-kink BPS solution exists, we expect that there is a repulsive force between the two kinks. This can be described by a superpotential, ${\cal W}(Y)$. But the low-energy theory of the kinks should be invariant under the $U(1)_R$ symmetry. This means that the superpotential must have $U(1)_R$ charge $+2$.

\para
We can determine the transformation of $Y$ under $U(1)_R$ by returning to the definition of $\sigma$ given in  \eqn{sigma}. Suppose that the first kink interpolates from $r=\kappa_{n-1}$ to $r=\kappa_{n}$, while the second interpolates from $r=\kappa_{n}$ to $r=\kappa_{n+1}$. This means that the value of $Q$ in \eqn{sigma} differs for each of the kinks: $Q^{2} = Q^{1} +2$, and the holomorphic coordinate can be written as
\be Y = \frac{v}{2}(X^{(1)}-X^{(2)}) - \alpha\nn\ee
But we saw earlier that the sigma-model field $r^1+ir^2=\rho e^{i\alpha}$ carries charge $+2$ under $U(1)_R$. This means that the unique superpotential allowed by the symmetries of the theory is 
\be {\cal W} \sim e^{-Y}\label{answer}\ee
As described in the introduction, this is the same superpotential that arises in the low-energy effective dynamics of ${\cal N}=2$ 3d gauge theory with $U(2)$ gauge group. In that case, the superpotential is a quantum effect, generated by instantons \cite{ahw}. Here we see that the same superpotential arising from the classical interaction between  kinks.

\subsection{The Force Between Kinks}

\para
The discussion given above was rather indirect. For this reason, we now present  a more explicit computation of the repulsive force between a pair of kinks. We use  a method which was developed in \cite{mantonkink} and previously applied to kinks in the multi-centered Taub-NUT metric \eqn{metric} in \cite{mmotw}. The basic idea is straightforward: you first construct the solutions for  individual domain walls and then build a new field configuration by superposing the two domain walls, separated by a large distance. As we will see, care must be taken about how these two solutions are patched in the middle. The resulting configuration of two domain walls is not a solution. If you let it evolve in time, the kinks will move apart. By computing their acceleration, you can extract the force experienced by the kinks, and hence the potential governing their relative separation.

\para
The full kink solutions involve both the hypermultiplet fields and the fields $\Phi$, $\Phi'$ and ${\cal B}$, ${\cal B}'$ on the D5-branes. The latter solve the familiar monopole Bogomolnyi equations \eqn{mono}. But it is known  that there is no force between BPS monopoles \cite{force}, with the repulsive magnetic force cancelled by the attractive force mediated by the scalar field. We expect that this cancellation continues to hold in the present context and that the only contribution to the force comes from the hypermultiplet fields. Unfortunately, we have been unable to prove this. We can, however, compute the force in the $g^2\rightarrow 0$ limit in which the gauge fields decouple and find that this agrees with the expectation \eqn{answer}.

\para
With these caveats, we focus only on the domain wall BPS equations \eqn{dw}. The force between two kinks then  reduces to the calculation of \cite{mmotw}.  We sketch this calculation below, and refer the reader to \cite{mmotw} for full details. 
We will take the first kink to interpolate from $r= \kappa_{n-1} $ to $r=\kappa_n$ and write the solution to \eqn{dw} as $r_{(1)}(x)$. This means that if we look far to the right, at $x-X^{(1)} \rightarrow \infty$, the kink has profile decaying as
\be \mbox{Kink 1:}\ \ \ \ \ r_{(1)} \rightarrow \kappa_n - 2\pi e^{- v(x-X^{(1)})}\nn\ee
Meanwhile, kink 2 interpolates from $r=\kappa_n$ to $r=\kappa_{n+1}$. We write the solution as $r_{(2)}(x)$. Far to the left of this kink, at $x-X^{(2)}\rightarrow -\infty$, the profile decays as
\be  \mbox{Kink 2:}\ \ \ \ \ r_{(2)} \rightarrow \kappa_n + 2\pi e^{+v(x-X^{(2)})}\nn\ee
We would like to patch these two solutions  to construct a configuration that looks like two far-separated kinks. Some care has to be taken in doing this because the point $r=\kappa_n$ where they join is a coordinate singularity in the target space metric \eqn{metric}. One consequence of this is that a naive field configuration, constructed by $r=r_{(1)} + r_{(2)} - \kappa_n$, will be genuinely singular at this point. To avoid this, we must first perform a field redefinition to a basis with canonical kinetic term at the point $r=\kappa_n$ where we wish to patch the configurations together. We can do this by writing
\be f[r-\kappa_n] = f[r_{(1)}-\kappa_n]+ f[r_{(2)}-\kappa_n]\nn\ee
where the function $f[r-\kappa_n]$ is given by
\be f[r-\kappa_n] = \sign(r-\kappa_n)\,\sqrt{|r-\kappa_n|}\nn\ee
In $r(x)$, we have constructed a static field configuration that interpolates between $r=\kappa_{n-1}$ and $r=\kappa_{n+1}$. Now we watch it evolve. The momentum $P$ of the first kink can be identified with the field momentum, integrated from $x=-\infty$ to some point $x=a$ which lies between the two kinks. The relevant contribution from the hypermultiplet field is
\be P = -\frac{1}{g^2}\int_{-\infty}^a  dx\ V(r)\,\partial_t r \partial_x r\nn\ee
Using the equations of motion, the acceleration of the kink is found to be a total derivative, given by
\be \dot{P} =\frac{1}{g^2}\left[ -\frac{1}{2}V(r)\left( (\partial_t r)^2 + (\partial_x r)^2\right) + \frac{1}{2}(\Phi-\Phi')^2 V^{-1}(r)\right]_{-\infty}^a\nn\ee
To compute the acceleration we used the truncated equations of motion without the gauge fields and with the scalars $\Phi,\,\Phi'$ set to their asymptotic values. 
Since the gauge fields and the scalars $\Phi,\,\Phi'$ solve the BPS monopole equations we do not expect them to contribute to the force between kinks.
Evaluating this on our two-kink configuration, the leading order contribution to the acceleration is independent of the choice of position $a$, as long as $X^{(1)}\ll a\ll X^{(2)}$. One finds
\be\dot{P} = -T v\, e^{-v (X^{(2)}-X^{(1)} )}\nn\ee
Armed with the knowledge of this acceleration, we can write down an effective action for the relative separation of the kinks defined in \eqn{y}. Taking the kinetic terms from \eqn{onekink}, the acceleration $\dot{P}$ is reproduced by the effective action
\be S_{\rm relative} = \int d^3x\ \frac{T}{v^2}\left( \partial \bar{Y}\partial Y -  v^2  |e^{-Y}|^2 \right)\label{kinkact}\ee
Such an action with ${\cal N}=2$ supersymmetry arises from superpotential
\be {\cal W} = \frac{T}{v}\, e^{-Y}\nn\ee
in agreement with the expectations from the symmetry analysis described above.

\subsection{Generalisations}

\begin{figure}[!h]
\begin{center}
\includegraphics[height=1.6in]{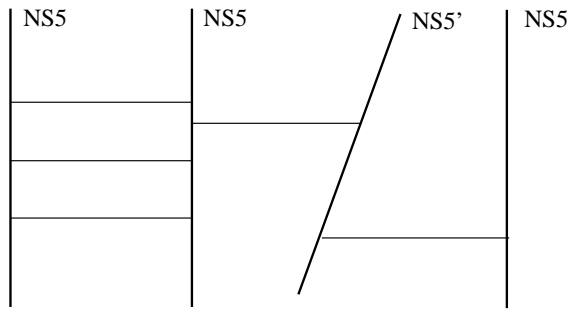}
\end{center}
\caption{A linear ${\cal N}=2$ quiver gauge theory.}\label{exfig}
\end{figure}

Consider a row of  NS5 and NS5$'$ branes with $n_a$ D3-brane suspended between the $a^{\rm th}$ and $(a+1)^{\rm th}$ brane. An example is shown in Figure \ref{exfig}. This results in a $d=2+1$ dimensional  ${\cal N}=2$ gauge theory with a linear quiver gauge group  $\prod_a U(n_a)$. There are  bi-fundamental hypermultiplets charged under adjacent gauge groups. If the D3-branes are suspended between parallel NS5-branes then the associated gauge group has an extra, adjoint chiral multiplet.

\para
After performing an S-duality, we can also write down the gauge group on the D5 and D5$'$ branes. Suppose,  starting from the left, that there are $K_1$ adjacent NS5-branes followed by $K_2$ adjacent NS5$'$-branes, followed by $K_3$ adjacent NS5-branes and so on. The D5-brane gauge theory has eight supercharges and  $\prod_iU(K_i)$ gauge group. There are defect hypermultiplets charged under the bi-fundamental representation of adjacent groups.These defect hypermultiplets should have non-canonical kinetic terms; this would be a non-Abelian extension of the proposal of \cite{mps}.

\para
D3-branes stretched between 5-branes of the same type appear as monopoles; D3-branes stretched between D5 and D5$'$ branes appear as kinks. The results of \cite{hw}, combined with those presented here, suggest that the low-energy dynamics of the $d=2+1$ dimensional gauge theory theory is captured by the classical dynamics of these interacting monopoles and kinks\footnote{We thank Ofer Aharony for helpful discussions and suggestions on this generalisation.}.

\para
There is currently little understanding of how such solitons of different types interact. (It may be that the recent work \cite{aki} studying the different boundary conditions of D3-branes in situations like this is of use here). However, we can glean some information by using this conjectured correspondence in reverse.

\EPSFIGURE{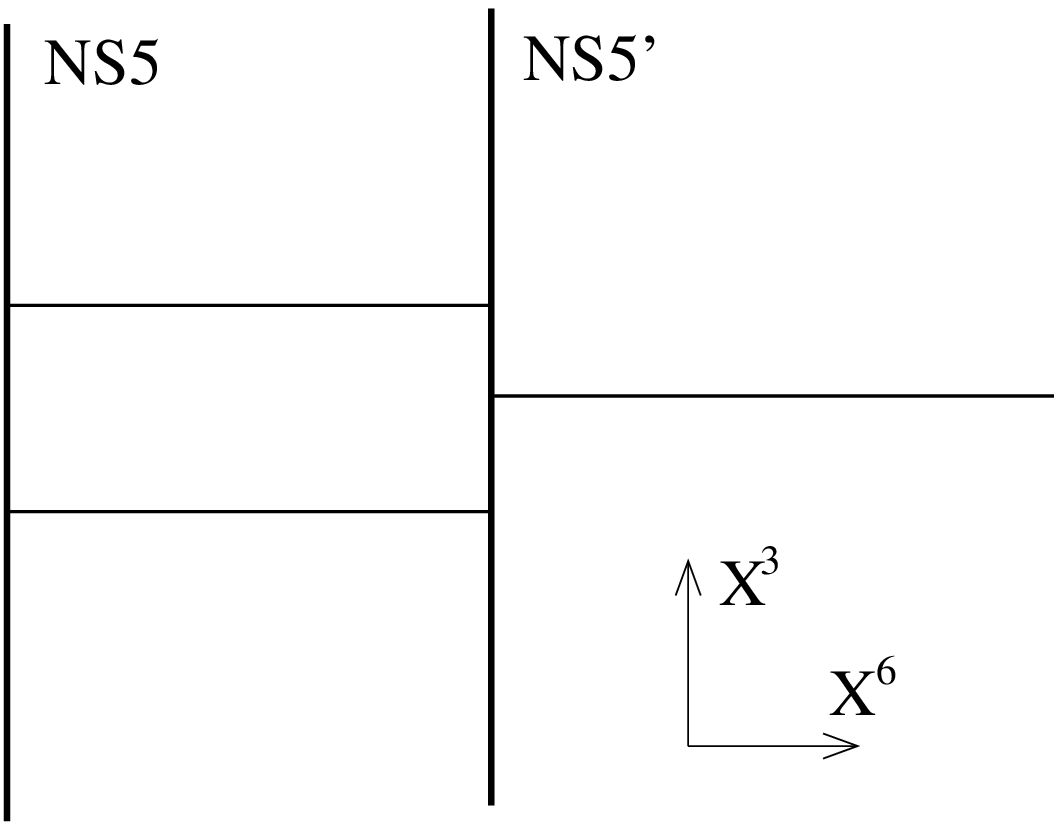,width=130pt}{}
\para
Consider the brane configuration shown in Figure 3. There are two D3-brane suspended between an NS5 and NS5$'$-brane. The difference with the set-up discussed in Section \ref{multikink} is the existence of a semi-infinity D3-brane stretched to the right. After an S-duality, this appears as a singular monopole or, equivalently, the insertion of a 't Hooft operator, in the theory on the D5$'$-brane. The dynamics of monopoles in the presence of 't Hooft operator has been studied in \cite{cherkis,cherkdur,moore}. Here we are interested in the dynamics of kinks in the background of a 't Hooft operator.

\para
The theory on the D3-branes is ${\cal N}=2$ $U(2)$ gauge theory coupled to a single fundamental flavour. The solution to the low-energy dynamics tells us that the Coulomb branch survives provided that the real mass for the flavour is not too large \cite{ahiss}. Translated to the brane picture, it means that there should be no force between the two mobile D3-branes provided that the semi-infinite D3-brane lies between them. In other words, the presence of a monopole lying between two kinks should be enough to ensure  that the kinks feel no force.

\para
Needless to say, it would be interesting to confirm this from a direct analysis of the moduli space of kinks and monopoles. For now, we merely mention that similar behaviour has been seen before in the study of domain walls in non-Abelian gauge theories \cite{nitta1,sakaime,amime}. In this case, the domain wall equations are very closely related to \eqn{dw}  but, in contrast to the theories studied here, moduli spaces of multi-domain wall solutions do exist provided that two domain walls in one part of the gauge group have a domain wall from a different part of  the gauge group lying between them. The result is that the domain wall moduli space exists only if the domain walls are interlaced 
in a similar manner to the monopole and kinks described above.

\section{Branes and Sigma-Model Lumps}\label{d1d5sec}

In this final section of the paper, we turn to a slightly different topic. We will see another feature of D-braneology that follows naturally from the proposal of \cite{mps}. 

\para
This time our interest lies in the system of D-branes with worldvolumes spanning
\be k\times D1&:& \ \ 01 \nn\\ D5\, &:&\ \ 012345 \nn\\ D5'&:&\ \ 012378\nn\ee
We place the all D5-branes coincident in the $x^6$ and $x^9$ directions. 

\para
We know that when there are multiple D5-branes, the D1-branes  can absorbed into their worldvolume where they appear as instantons. But what happens in the case of a single D5 and D5${}^\prime$ brane? Now there is no gauge symmetry enhancement on their overlap. But, nonetheless, when the two branes are coincident, it is natural to expect that D-strings can be absorbed into their worldvolume.  

\para
To see that this expectation is correct, we can look at the worldvolume theory of $k$ D1-branes.  This is a $d=1+1$ $U(k)$ gauge theory with ${\cal N}=(2,2)$ supersymmetry. The vector multiplet contains a complex scalar $\Sigma = X^6+iX^9$. There are a further 3 adjoint chiral multiplets, with complex scalars
\be \ \ \ X = X^2+iX^3\ \ \ ,\ \ \  Z=X^4+iX^5\ \ \ ,\ \ \ W=X^7+iX^8
\nn\ee
The D1-D5 strings give rise to two chiral multiplets, $Q$ and $\tilde{Q}$.  The D1-D5$'$ strings give rise to two further chiral multiplets that we call $P$ and $\tilde{P}$. Both $Q$ and $P$ transform in the fundamental of the gauge group; $\tilde{Q}$ and $\tilde{P}$ transform in the anti-fundamental.

\para
The theory on the D-strings is an obvious generalisation of the usual D1-D5 system. The theory has the usual Coulomb branch in which $Q=\tilde{Q}=P=\tilde{P}=0$, while the adjoint scalars mutually commute. More interesting is the Higgs branch of the theory, which occurs when $\Sigma = Z= W=0$. The remaining fields are constrained to obey
\be V &=& {\rm Tr}\,\left(QQ^\dagger - \tilde{Q}^\dagger \tilde{Q}+ PP^\dagger - \tilde{P}^\dagger \tilde{P} + [X,X^\dagger] \right)^2 + {\rm Tr}\,|Q\tilde{Q}|^2 + {\rm Tr}\,|P\tilde{P}|^2=0\nn\ee
%
%
%
%
One solution arises if we set $\tilde{Q} = P=0$. Then $Q$ and $\tilde{P}$ must obey the D-term condition
\be QQ^\dagger - \tilde{P}\tilde{P}^\dagger + [X,X^\dagger] =0 \label{higgs2}\ee
After dividing through by the $U(k)$ gauge symmetry, we're left with a Higgs branch of real dimension $4k$. This describes the D-strings absorbed in the worldvolume of the D5-branes. 

\para
However, there is also a second branch  that arises by setting $Q=\tilde{P}=0$, with $\tilde{Q}$ and $P$ subject to
\be -\tilde{Q}^\dagger\tilde{Q} + PP^\dagger+ [X,X^\dagger] = 0\label{higgs1}\ee
This again has dimension $4k$. The two branches meet at the singular point $Q=\tilde{Q} = P=\tilde{P}=0$.

\para
We now turn to the perspective of the D5-branes, described by the theory \eqn{lag} and \eqn{lhyper}. With the D5-brane coincident, we can set $\Phi=\Phi'=0$ and focus on the hypermultiplet and gauge fields alone. We expect that the D-strings should arise as BPS solitons in the D5-brane theory. Indeed, such solitons exist: they are sigma-model lumps, arising from $\Pi_2$ of the target space. For the multi-centred Taub-NUT space \eqn{metric}, we choose a ground state, say $r=\kappa_n$.  There are then two 2-cycles in the geometry which intersect at this point. One of these is the 2-cycle with $r=\kappa_{n-1}$ and $r=\kappa_n$ at the poles; the other has $r=\kappa_n$ and $r=\kappa_{n+1}$ at the poles. We may choose to wrap either of these 2-cycles, resulting in a soliton that is localised in the $(x^2,x^3)$ plane and extended along the $x^1$ direction. The tension of the soliton is determined by the Kahler class of the target space. With the choice of centres given by \eqn{centres}, this is 
\be T = \frac{8\pi^2}{g^2}\nn\ee
This is the same as the tension of the D-string.

\para
We can build a charge $k$ BPS soliton by wrapping either the first 2-cycle $k$ times or the second 2-cycle $k$ times (but not by wrapping both). For either choice, the moduli space is essentially that of $k$ ${\bf CP}^1$ lumps. (If $c\neq 0$ in \eqn{v} then the ${\bf CP}^1$ is squashed but the K\"ahler class remains the same). The moduli space of $k$ ${\bf CP}^1$ lumps is known to have real dimension $4k$ \cite{ward}, in agreement with the dimension of the Higgs branch computed above. 

\para
In fact, we can do better in relating the soliton moduli space to the Higgs branch \eqn{higgs1}. A D-brane construction of sigma-model lumps was previously presented in \cite{vib}. In that paper, the gauged linear sigma model describing the lump moduli space was given by an ${\cal N}=(2,2)$ $U(k)$ gauge theory with a single adjoint chiral multiplet $X$, a fundamental chiral multiplet $Q$ and an anti-fundamental $\tilde{P}$. This is not the same as the gauge theory on the D-string, but the resulting Higgs branches is again given by \eqn{higgs1}. (We note that the Higgs branch is thought to capture the topology and K\"ahler class of the soliton moduli space. But the metrics do not match. In particular, the metric on the moduli space of ${\bf CP}^1$ lumps is known to suffer from logarithmic divergence. See, for example, \cite{intseib} for a recent discussion of this issue.)

\para
Finally, it is natural to identify the two Higgs branches \eqn{higgs1} and \eqn{higgs2} with the moduli space of lumps arising from the two intersecting 2-cycles.

\appendix


\section{A Gauged Linear Model for the Hypermultiplet}\label{glsmapp}

The non-linear sigma model described in Section \ref{intbsec} suffers from a number of coordinate artefacts, most prominent among them the lack of continuity of the D-term. In this appendix, we present an equivalent description in terms of a gauged linear model, first  introduced in \cite{dm}.

\para
In the gauged linear description, the dynamical degrees of freedom consist of $N$ hypermultiplets, with bosonic fields $(q_i,\tilde{q}_i)$, $i=1,\ldots ,N$, subject to $N-1$ constraints. These constraints are imposed by introducing $N$ vector multiplets with gauge fields $C_i$ and complex scalars $\varphi_i$. (The sum of these vector multiplets will decouple so only $N-1$ constraints remain). In this language, the hypermultiplet Lagrangian in \eqn{lag} is
\be {\cal L}_{\rm hyper} &=& \sum_{i=1}^N \Bigg\{ |{\cal D}_\mu q_i|^2 + |{\cal D}_\mu\tilde{q}_i|^2  + \frac{1}{4e_i^2} |dC_i|^2 + \frac{1}{2e_i^2}|\partial_\mu \varphi_i|^2
\label{glsm}\\&&\ \ \ \ \ \ \ \ \  + (\varphi_i-\varphi_{i+1} +\Phi-\Phi')^2(|q_i|^2 + |\tilde{q}_i|^2) 
 \nn\\ &&\ \ \ \ \ \ \ \ \ + e_i^2 \left|\tilde{q}_iq_i- \tilde{q}_{i-1}q_{i-1}\right|^2 + \frac{e_i^2}{2}\left(|q_i|^2 - |\tilde{q}_i|^2 - |q_{i-1}|^2 + |\tilde{q}_{i-1}|^2\ -\zeta_i\right)^2  
\Bigg\}\nn\ee
Here the field $q_i$ has charge $(+1,-1)$ under $U(1)_i\times U(1)_{i+1}$ while $\tilde{q}_i$ has the opposite charge. The covariant derivatives are therefore ${\cal D}q_i = \partial q_i - i C_iq_i + i C_{i+1}q_i$ and ${\cal D}\tilde{q}_i = \partial\tilde{q}_i + i C_i\tilde{q}_i - i C_{i+1}\tilde{q}_i$. 
Ultimately, we will send $e_i^2\rightarrow \infty$, so that the gauge fields can be integrated out while  two of the potential terms above are imposed as constraints. However, for now it will prove useful to leave $e_i$ finite. 

\para
The FI parameters $\zeta_i$ in \eqn{glsm} must satisfy the requirement that $\sum_{i=1}^N\zeta_i=0$. This is because nothing is charged under the overall gauge group $\oplus_i U(1)_i$. We will shortly relate them to the centres $\kappa_i$ in the Gibbons-Hawking metric \eqn{v}. 

\para
We still have to specify the D and F-term couplings to the D5-brane gauge fields in \eqn{lag}. These are particularly simple in the gauged linear approach. The D5-brane gauge field $A_\mu$ couples to each $q_i$ with charge $+1$ and to each $\tilde{q}_i$ with charge $-1$. The corresponding D and F-terms are
\be D =  \frac{1}{2}\sum_{i=1}^N |q_i|^2 - |\tilde{q}_i|^2\ \ \ \ {\rm and}\ \ \ \ F=\frac{1}{\sqrt{2}}\sum_{i=1}^N \tilde{q}_iq_i\label{glsmdf}\ee

\para
When the D5-branes are coincident, so $\Phi-\Phi'=0$, the moduli space of vacua of \eqn{glsm} arises by imposing the vanishing of the terms in the final line and dividing out by gauge transformations. The result is the Gibbons-Hawking metric \cite{dm}. The function $V$ in \eqn{v} has  $c=0$; it is an ALE metric rather than an ALF metric\footnote{There is a hyperK\"ahler quotient construction of the ALF metrics with $c\neq 0$ \cite{gr} and this has found a number of applications in gauge theories \cite{ms,ns,witten}. Because the coefficient $c$ does not affect the force between kinks computed in Section \ref{multikink},  we choose to  work here with the simpler model with $c=0$. }. The explicit relationship between the FI parameters $\zeta_i$ and the centres $\vec{\kappa}_i = (0,0,\kappa_i)$ depends rather critically on a choice of ordering and minus signs. In the geometry, we pick $\kappa_{i+1} > \kappa_i$. In the gauge theory, we will require that $\zeta_i > 0$ for $i=2,\ldots,N$. This means that we must have $\zeta_1 = -\sum_{i=2}^N \zeta_i$. Then one can show that
\be \zeta_i = \frac{1}{N} \left(\kappa_{i}- \kappa_{i-1}\right) \ \ \ \ \ i=2,\ldots,N\nn\ee
When the D5-branes are separated, we have $\Phi-\Phi'\neq 0$. Now the moduli space is lifted, leaving behind  $N$ isolated vacua. These are characterised by 
\be \mbox{\underline{Vacuum $k$:}\ \ }  &&q_i=0 \ \ \ \  i=1\ldots k\label{glsmvev}\\ && \tilde{q}_i=0\, \ \ \ i=k,\ldots N\nn\\ &&   \varphi_i-\varphi_{i+1} = \Phi-\Phi'\ \ \ \  \ i\neq k\nn\ee
Note that in vacuum $k$ we have $q_k=\tilde{q}_k=0$. All other hypermultiplets, $i\neq k$, have either $q_i=0$ or $\tilde{q}_i=0$. The expectation values of these fields are easily determined by setting the potential equal to zero in  \eqn{glsm}.

\subsubsection*{Kinks in the Gauged Linear Model}

We now repeat the analysis  of Section \ref{kinksec} using the notation of the gauged linear model. BPS domain walls in these models were first discussed in \cite{mewall}. Here, again, the novel ingredient is the coupling to the D5-brane fields, wherein the kink looks like a magnetic monopole. 

\para
Here we will be interested in the situation where the kink interpolates between vacuum $k$ and vacuum $k+1$. For such kinks, the vacuum expectation values \eqn{glsmvev} tells us that, throughout the kink profile,  either $q_i=0$ or $\tilde{q}_i=0$ for all $i=1,\ldots,N$. This means that $\tilde{q}_iq_i=0$ for each $i$ so that the F-term in \eqn{glsmdf} vanishes. We can choose a gauge such that $C_i=0$. 
The remaining terms can be rearranged to write the tension of the kink as 
\be T &=& \frac{1}{g^2} \int dx^3 dz d\bar{z}\ \Bigg\{ \frac{1}{2}\left(\partial_3\Phi - (F_{45} - \sum_{i=1}^N(|q_i|^2-|\tilde{q}_i|^2)\,\delta^2(z,\bar{z})\right)^2 +\frac{1}{2} |\partial_z\Phi +i F_{3z}|^2 \nn\\  &&\ \ \ \ \ \ \ \ \ \ \ \ \ \ \ \ \ \ \ \ \ \ \ \ \ \ \ \  -\ \sum_{i=1}^N(|q_i|^2-|\tilde{q}_i|^2)\,\partial_3\Phi\,\delta^2(z,\bar{z}) 
 +\frac{1}{2}\epsilon_{ijk}F_{ij}\partial_k\Phi\Bigg\}
\nn\\ && \!\!\!\!\!\!\!\!\!\!+\ \frac{1}{g^2}\int dx^3 dw d\bar{w}\ \Bigg\{ \frac{1}{2}\left(\partial_3\Phi' - (F'_{78} + \sum_{i=1}^N(|q_i|^2-|\tilde{q}_i|^2)\delta^2(w,\bar{w})\right)^2  + \frac{1}{2}|\partial_w\Phi' +i F'_{3w}|^2
\nn\\ &&\ \ \ \ \ \ \ \ \ \ \ \ \ \ \ \ \ \ \ \ \ \ \ \ \ \ \ \ \ +\sum_{i=1}^N(|q_i|^2-|\tilde{q}_i|^2)\,\partial_3\Phi'\,\delta^2(w,\bar{w}) + \frac{1}{2}\epsilon_{ijk}F'_{ij}\partial_k\Phi'\Bigg\}  \nn\\
&&  \!\!\!\!\!\!\!\!\!\! +\  \int dx^3 \ \sum_{i=1}^N\Bigg\{ |\partial_3q_i + (\varphi_i-\varphi_{i+1}+\Phi-\Phi')q_i|^2-  |\partial_3\tilde{q}_i - (\varphi_i-\varphi_{i+1}+\Phi-\Phi')\tilde{q}_i|^2 \nn\\ &&\ \ \ \ \ \ \ \ \ \ \ \ \ \ \ \ \ \    +\ \frac{1}{2e_i^2}\left(\partial_3\varphi_i -  e_i^2(|q_i|^2 - |\tilde{q}_i|^2 - |q_{i-1}|^2 + |\tilde{q}_{i-1}|^2-\zeta_i\right)^2
\nn\\ &&\ \ \ \ \ \ \ \ \ \ \ \ \ \ \ \ \ \  - \ \partial_3\left((\varphi_i-\varphi_{i+1})(|q_i|^2 -|\tilde{q}_i|^2)\right)-(\Phi-\Phi')\partial_3(|q_i|^2 - |\tilde{q}_i|^2) + \zeta_i\partial_3\varphi_i\Bigg\}
\nn\ee
%
Here $x^i$ refers to $x^{3,4,5}$ on the D5-brane and $x^{3,7,8}$ on the D5$'$-brane. The BPS equations can again be found nestling inside the total squares. Setting them equal to zero gives the energy of the BPS kink, 
\be
T =\frac{1}{g^2} \sum_{i=1}^N \zeta_i\Delta\varphi_i + \frac{1}{2g^2}\int dx^3dzd\bar{z} \ B_i\partial_i\Phi + \frac{1}{2g^2}\int dx^3dwd\bar{w}\ B'_i\partial_i\Phi'    
\nn\ee
where we have used the fact that the vacuum conditions \eqn{glsmvev} hold as $x^3\rightarrow \pm\infty$. As in Section \ref{kinksec}, the fields $B_i$ and $B'_i$ carry no magnetic flux, ensuring that the last two terms vanish. Once again we find that BPS kinks have tension $T=4\pi v/g^2$.

\newpage
\section*{Acknowledgements}

We are grateful to Joe Polchinski for useful correspondence and to Ofer Aharony and Micha Berkooz for useful discussions. We are supported by STFC and by the European Research Council under the European Union's Seventh Framework Programme (FP7/2007-2013), ERC Grant agreement STG 279943, Strongly Coupled Systems

\end{document}